\documentclass[prd,letterpaper,tightenlines,
showpacs,preprintnumbers,nofootinbib,floatfix,psfig]{revtex4}

\usepackage{amsmath,amssymb,amsfonts,graphicx,pifont,dcolumn}
\usepackage{color}
\usepackage{cancel}
\usepackage{ulem}
\RequirePackage{slashed}

\begin{document}

\title{Comments on glueballs as gravitons }

\author {Matteo Rinaldi}

\author{ Vicente Vento }

\affiliation{Departamento de F\'{\i}sica Te\'orica-IFIC, Universidad de Valencia- CSIC, 46100 Burjassot (Valencia), Spain.}

\date{\today}

\begin{abstract}

The graviton solutions for the glueball spectrum of ref.~\cite{Rinaldi:2017wdn} interpreted in a different manner lead to very interesting results which we describe in this comment.

\end{abstract}


\maketitle

\section{Introduction}
In previous work we studied in detail the graviton solutions for different models of $AdS_5$ and established a detailed comparison with the lattice glueball spectrum \cite{Rinaldi:2017wdn}. In here we recall the same equations and models for the graviton and proceed inversely, we plot the solutions of the $AdS_5$ modes and over them we seed the lattice data. To go from the $AdS/QCD$ solutions to the lattice data we only use one scale. This way of proceeding leads to very interesting comparison which merits the present comment.

Let us show the precise lattice  data with their corresponding errors, which have been obtained from the mentioned calculations by summing all different types of errors in quadrature.

\begin{table} [htb]
\begin{center}
\begin{tabular} {|c c c c c c c|}
\hline
& $0^{++}$&$2^{++}$&$0^{++}$&$2^{++}$&$0^{++}$&$0^{++}$\\
\hline
MP & $1730 \pm 94$ & $2400 \pm122$ & $2670 \pm 222 $&  & &  \\
\hline
YC & $1719 \pm 94$ & $2390 \pm124$ &  &  &  &  \\
\hline
LTW & $1475 \pm 72$ & $2150 \pm 104$ & $2755 \pm 124$& $2880 \pm 164 $& $3370 \pm 180$& $3990 \pm 277$  \\
\hline
Lattice & $1631\pm 50$ & $2313 \pm 68$ & $2713 \pm 127 $ &$2880 \pm164$ & $3370 \pm 180 $ &$ 3990 \pm277$ \\
\hline
\end{tabular}
\caption{Glueball masses (MeV) from lattice calculations MP \cite{Morningstar:1999rf}, YC \cite{Chen:2005mg} and LTW \cite{Lucini:2004my} and Lattice (average)}
\label{masses}
\end{center}
\end{table}

We have not included the lattice results from the unquenched calculation \cite{Gregory:2012hu} to be consistent, which however,  in this range of masses and for these quantum numbers are in agreement with the shown results within errors.

\begin{figure}[htb]
\begin{center}
\includegraphics[scale= 0.6]{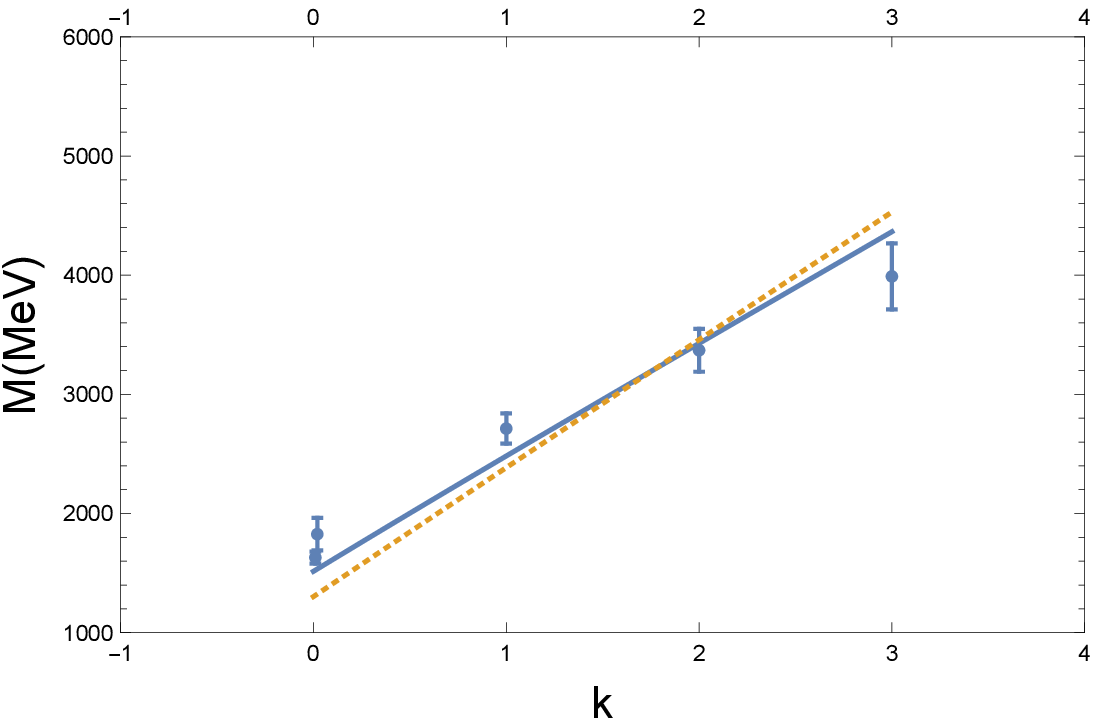} 
\hskip 0.5cm 
\includegraphics[scale= 0.6]{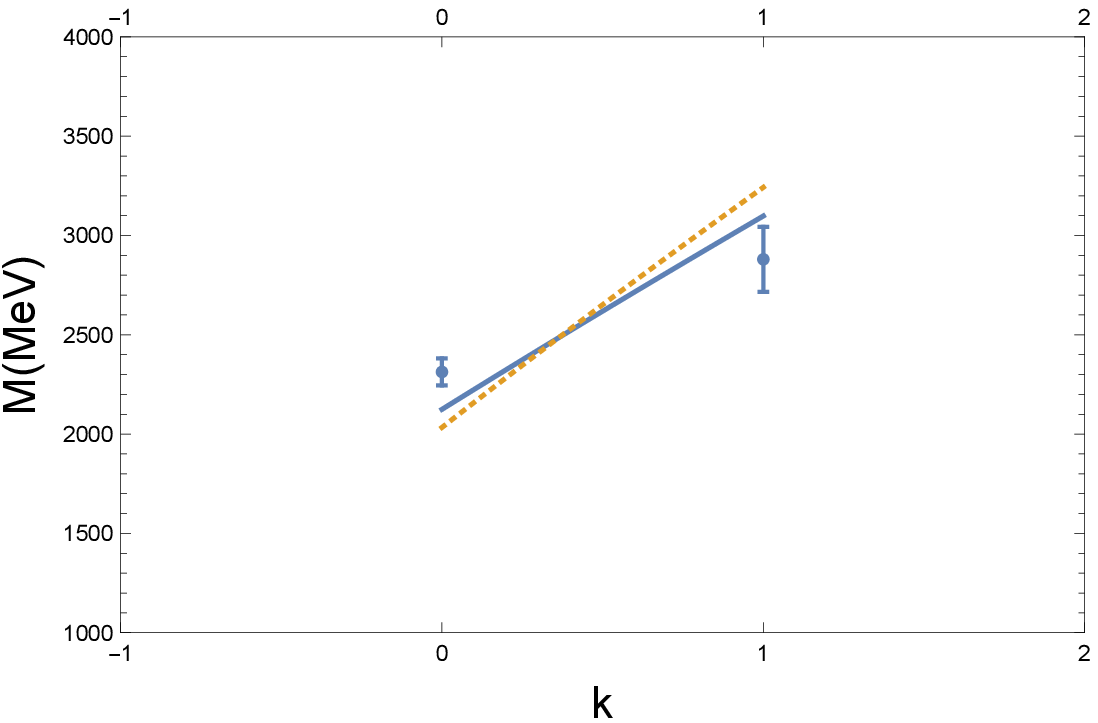} 
\end{center}
\caption{Gluebal spectrum obtined by the hardwall approach. The figure on the right shows the $0^{++} $ glueballs. The figure on the left shows the tensor $2^{++}$ glueballs. The solid line shows the solution for Cauchy  boundary conditions and the dotted line for Neumann boundary conditions.} 
\label{hardwall}
\end{figure}

\section{Glueballs as  gravitons}

The first calculation discussed in ref. \cite{Rinaldi:2017wdn} is the hardwall calculation \cite{BoschiFilho:2002vd}. We show in Fig.\ref{hardwall} two interpolations of the graviton modes one for Cauchy and the other for Neumann boundary conditions as a function of the mode number for scalar and tensor glueballs. Our notation for $k= 0,1,2,\ldots$ differs from that of ref. \cite{BoschiFilho:2002vd}  $k = k_B -1$.The energy scale, the only free parameter used, has been fixed to optimize the agreement. The fit comes out  linear and the slope is fixed by $AdS$.  We could say the agreement is fair, but the slope of the lattice calculation and $AdS$ are certainly different.

\begin{figure}[htb]
\begin{center}
\includegraphics[scale= 0.6]{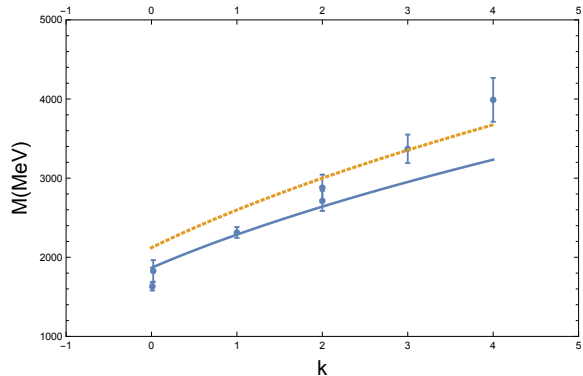} 
\end{center}
\caption{Gluebal spectrum obtained from the graviton of the softwall model of ref.\cite{Colangelo:2007pt}. The solid curve fits the lower scalar glueball masses and the dashed curve the higher glueball scalar masses.} 
\label{softwallColangelo}
\end{figure}

The second model we analyzed was that of ref.\cite{Colangelo:2007pt}. The mode equations for the scalar and tensor components of the graviton lead to the following spectrum

\begin{eqnarray}
M^2_0 & = & 4k+ 8, \nonumber \\
M^2_2 &=  & 4k + 12, \nonumber
\end{eqnarray}
with $k=0,1,\ldots$. We can write a unique equation for both tensor and scalar graviton components assuming that the  tensor  graviton component start at $k=1$, while the scalar graviton component at $k=0$. The relation to the mode number is quadratic in the mass. We plot in Fig. \ref{softwallColangelo} this equation and find a good agreement with the data within errors. We show two fits, one aimed at fixing the lower mass glueballs and the other aimed at fitting the higher mass glueballs. The experimental value for the lowest scalar glueball  $1631 \pm 50$ GeV is too low, while its N infinity limit of ref.\cite{Lucini:2004my}   $1827 \pm 136$ is better reproduced. Despite the goodness of the fit, it seems that a linear relation is wanted by the data instead of a quadratic one. It must be recalled \cite{Rinaldi:2017wdn} that the scalar and tensor graviton components in this model satisfy the same equations as the scalar and tensor fields of ref. \cite{Colangelo:2007pt}.

\begin{figure}[htb]
\begin{center}
\includegraphics[scale= 0.6]{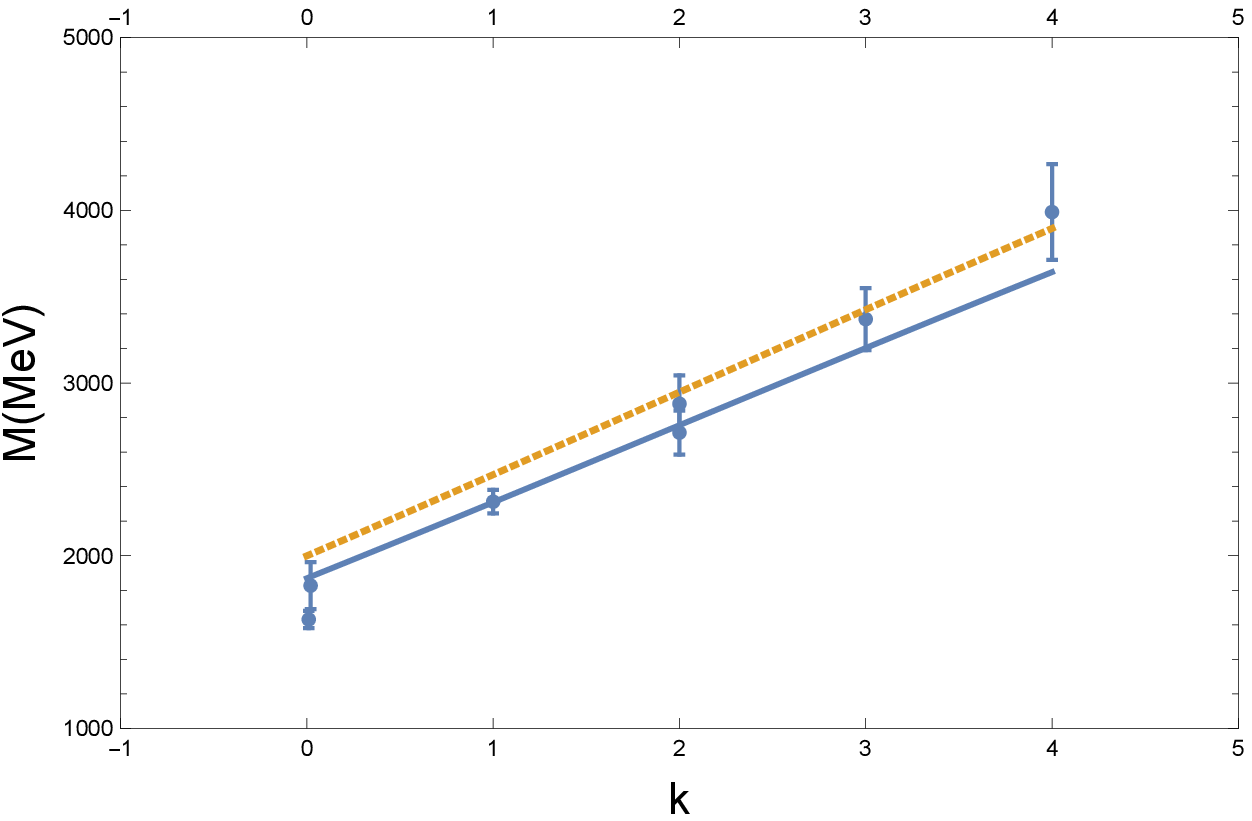} 
\hskip 0.5cm
\includegraphics[scale= 0.6]{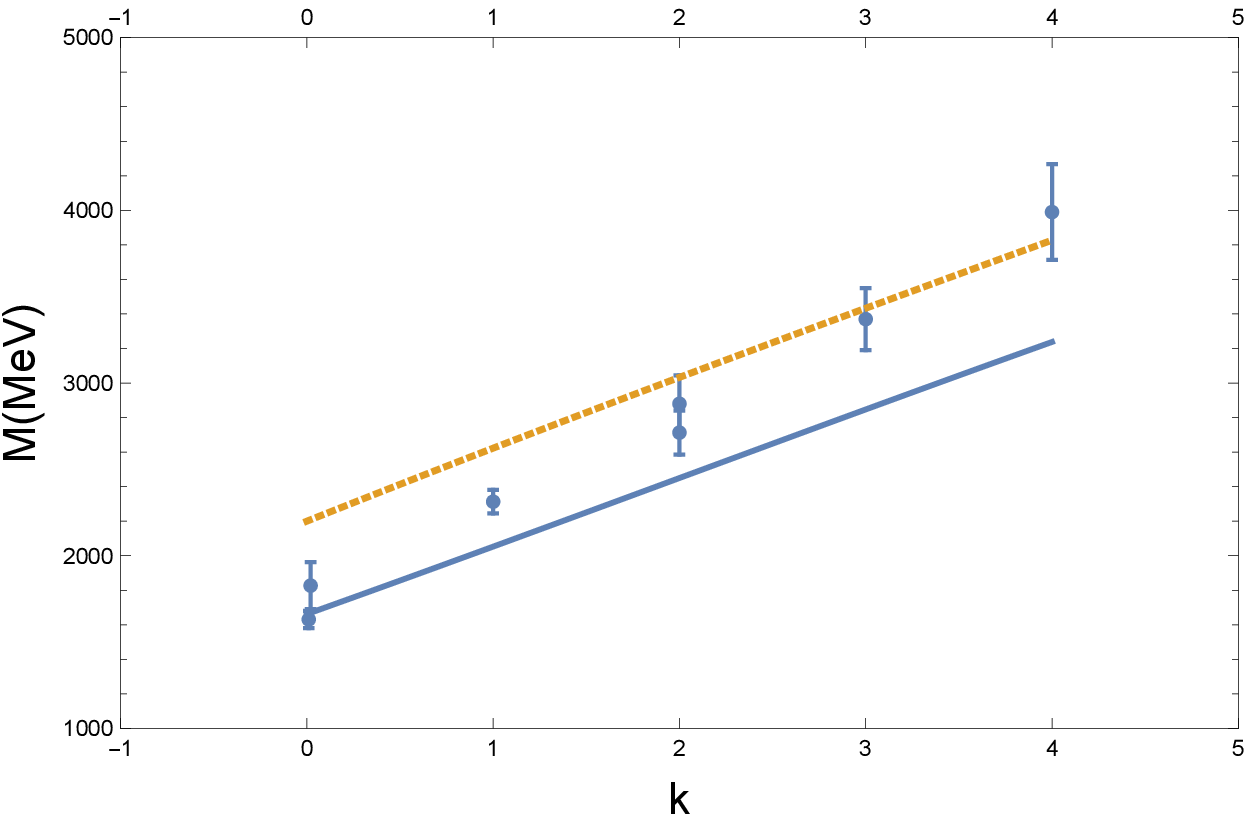} 
\caption{Left: We show the glueball spectrum obtained using the model of ref.\cite{Rinaldi:2017wdn}.  The solid curve is fitted to the lower glueball masses and agrees relatively well with the  $N\rightarrow \infty$ limit of the lowest scalar glueball of ref.\cite{Lucini:2004my}. The dotted curve aims its fitting the  the higher masses. Right: The glueball spectrum using the same model but adding the conventional mass term for the tensor component. The solid line aims at fitting the  lowest scalar massess and the dotted curve the tensor glueball masses.}
\label{softwallRinaldi}
\end{center}
\end{figure}

Finally we plot the results of the model of ref \cite{Rinaldi:2017wdn} in Fig. \ref{softwallRinaldi}. In the left figure we use the same equation for the scalar and the tensor component of the graviton, as comes out naturally from the graviton equations. The fits are good. The two curves aim at fitting the lower mass glueballs (solid) and the higher mass glueballs (dotted) respectively. If we compare this fit with the prevoius fit in Fig \ref{softwallColangelo} we see that the lattice data require a linear relation and the graviton in our simple model of $AdS$  gives a linear fit and a reasonable slope. In the right figure we add to the graviton a conventional mass term to its tensor component and the result is not so good, due to the fact that the scale factor of the scalar is too large for the tensor with mass. Thus it seems the graviton is approaching the glueball spectrum without the need of an additional mass term.

\section{Conclusion}
In our previous work \cite{Rinaldi:2017wdn} we have aimed at fitting the exact spectrum. This has produced some difficulties because as it is clearly seen in the present comment, there are states in the $AdS$ spectrum that are missing in the lattice spectrum. For example the scalar state for the $k=1$ mode is missing, and the same happens with the $k=3,4$ tensor modes. Our attitude here has been very different. We plot the dynamics of $AdS$ as it comes out and seed the lattice data, and they fall close to the dynamical curves. Why are some states missing? The missing tensor states correspond to high modes and lattice calculations could have missed them, however the scalar mode at $k=1$ should have been found with the actual level of precision. Thus we see two possibilities, either the state does not exists and the dynamics of $AdS$ corresponding to QCD is much more complicated then the one we have studied, or the state will be found and $AdS$ as used is really explaining the spectrum. The last possibility is exciting. In any case looking at Figs.\ref{softwallColangelo} and \ref{softwallRinaldi} one has to accept that $AdS$ is really telling us something about the strong interactions.

\section*{Acknowledgments}
We thank Sergio Scopetta, Tatiana Tarutina  and Marco Traini for discussions. 
This work was supported in part by Mineco under contract 
FPA2013-47443-C2-1-P, Mineco and UE Feder under contract FPA2016-77177-C2-1-P, GVA- PROMETEOII/2014/066 and SEV-2014-0398.

\end{document}